\documentclass[aps,preprint,showpacs,superscriptaddress,groupedaddress]{revtex4-1}  
\usepackage{graphicx}  
\usepackage{dcolumn}   
\usepackage{bm}        
\usepackage{amssymb}   
\usepackage{amsmath}   
\usepackage{listings}
\usepackage{array}
\usepackage{graphicx}
\usepackage{lineno}
\usepackage{dcolumn}
\usepackage{color}
\usepackage{overpic}
\usepackage{multirow}
\usepackage{mathtools}
\usepackage{hyperref}

\newcommand{\al}{\alpha}
\newcommand{\ep}{\epsilon}
\newcommand{\sg}{\sigma}
\newcommand{\vx}{\vec{x}}
\newcommand{\nbins}{N_{\text{bins}}}
\newcommand{\intaz}{\int_{-\infty}^{+\infty}}

\begin{document}

\widetext


\title{Understanding the boosted decision tree methods with the weak-learner approximation}
\author{Li-Gang Xia \\ Department of Physics, Warwick University, CV4 7AL, UK}

\begin{abstract}
   
   Two popular boosted decsion tree (BDT) methods, Adaptive BDT (AdaBDT) and Gradient BDT (GradBDT) are studied in the classification problem of separating signal from background assuming all trees are weak learners. The following results are obtained. a) The distribution of the BDT score is approximately Gaussian for both methods. b) With more trees in training, the distance of the expectaion of score distribution between signal and background is larger, but the variance of both distributions becomes greater at the same time. c) Extenstion of the boosting mechanism in AdaBDT to any loss function is possible. d) AdaBDT is shown to be equivalent to the GradBDT with 2 terminal nodes for a decision tree. In the field high energy physics, many applications persue the best statistical significance. We also show that the maximization of the statistical significance is closely related to the minimization of the loss function, which is the target of the BDT algorithms. 

\end{abstract}

\maketitle

\section{Introduction}

Machine learning (ML) is generally used in the field of data analysis. For example, it is used for particle identification~\cite{atlas_tauid,atlas_photonid, atlas_bjetid} and to search for rare signals~\cite{atlas_hbb,cms_hmumu} while suppressing the background as much as possible in high energy physics (HEP). Among various ML methods, boosted decision tree (BDT) methods are shown to be effective and robust in some cases~\cite{hjyang1,hjyang2}. Especially, a new BDT algorithm~\cite{QBDT} is developed to take into account systematical uncertainties in the training. All BDT methods use decison trees as weak learners and obtain strong classification power by combining many weak learners. The training process in any BDT algorithm is to minimize some kind of loss function in a stage-wise way. In this paper, two classic BDT algorithms, Adaptive BDT (AdaBDT) and Gradient BDT (GradBDT), are studied assuming all learners (decision trees) are weak enough. With this assumption, we are able to obtain the probability distribution function (PDF) of the BDT score, study its evolution with increasing number of trees, and investigate the potential relation between two methods.
Besides, for many applications in HEP, great efforts are made to improve the signal significance with the existence of hugh background. 
It is possible to study the relation between the minization of the loss function and the maximization of the signal significance under the weak-learner approximation. 

We will only focus on the application of the BDT method in the two-class classification problem, like signal and background categories in HEP. The basic elements for a BDT algorithm is reviewed in Sec.~\ref{sec:bdtelement}. The score PDF is derived for the AdaBDT method in Sec.~\ref{sec:AdaBDT} and for the GradBDT method in Sec.~\ref{sec:GradBDT}. The evolution the statistical significance in HEP with the number of trees is studied in Sec.~\ref{sec:HEP}. We will summarize the conclusions in Sec.~\ref{sec:summary}.

\section{Review of the basic elements in a BDT algorithm}~\label{sec:bdtelement}

In this section, we introduce the basic elements of a BDT algorithm. They are summarized in Table~\ref{tab:rev_bdt} for comparing the two BDT methods. We have two categories, signal and background, and a number of variables (denoted by a vector $\vx$) used for classification. For an unknown instance, a real number, namely the BDT score $y$, will be assigned according to the variables associated with this instance. Thus $y$ is a function of $\vx$. Every instance in the training has a true value $Y(\vx)$, which is 1 if it belongs to the signal category and -1 if it belongs to the background category by convention. Any BDT algorithm is trying to minimize a loss function, $L(y)$, which is a functional of the score function $y(\vx)$. For example, $L(y) = \sum_{\vx} \frac{1}{2}(y(\vx)-Y(\vx))^2$ or $L(y) = \sum_{\vx}e^{-y(\vx)Y(\vx)}$ are widely used.

\begin{table}
   \caption{\label{tab:rev_bdt}
	Summary of the basic elements for the two BDT methods. Here $m$ denotes a $m$-tree training. $\epsilon$ is the misidentification rate in the Adaptive BDT. $w$ is negative gradient of the loss function in the Gradient BDT. 
   }
   \begin{tabular} {llll}
	\hline\hline
	Quantity &  AdaBDT value & GradBDT value\\
	\hline
	Input variables &$\vec{x}=(x_1,x_2,\cdots)$& same  \\
	True value $Y(\vx)$ & -1, 1 & same  \\
	Tree output & $k(\vx)=$ -1, 1& negative gradient, $w(\vx)\propto -\frac{\partial L(y)}{\partial y}$   \\
	Tree weight & $\alpha = \frac{1}{2}\ln \frac{1-\epsilon}{\epsilon}$  & 1 \\
	Tree update & apply $e^{\alpha}$ to wrong guess & fit the residues \\ 
	Node split & Gini index reduction & loss function reduction\\ 
	BDT score $y_m$& $y_m = y_{m-1} + \alpha_m k_m $  & $y_m = y_{m-1} + w_m$ \\ 
	\hline
	Loss function $L( y_m)$ & $\sum_{\vec{x}} e^{-Y(\vec{x})y_m(\vec{x})}$& any differentiable loss function \\ 
	\hline\hline
   \end{tabular}
\end{table}

In practice, hundreds of trees are involved for a training and there is a number of terminal nodes per tree. Each tree has an output reflecting the classification result and a weight reflecting the confidence of the classification. Taking the AdaBDT as example, the output (denoted by $k(\vx)$) is 1 if an instance falls into the signal-dominated terminal node while -1 if it falls into a background-dominated terminal node. The tree weight $\alpha$ is a function of the misclassification rate $\epsilon$, namely, $\alpha = \frac{1}{2}\ln\frac{1-\epsilon}{\epsilon}$ where $\epsilon$ is defined as the fraction of misclassified instances. Basically, if a tree has a low misclassification rate, the probability of correct classification is high and this tree shall contribute more to the final output (BDT score).

The crucial element is the boosting mechanism. Subsequent tree will pay more attention to those instances misclassified by the current tree. For the AdaBDT, misclassified instances are assigned a heavier weight of $e^{\alpha}$ and subsequent tree will is builted with the weighted samples. The GradBDT works differently. It is to approach the true value in a stage-wise way. Starting from a random value like 0 for all instances, the first tree is trained to fit the difference between the true value and the initial guess, namely, $Y(\vx)-0$.  This is realized by searching for the output $w_1(\vx)$ to maximize $L(0) - L(w_1(x))$ so that the loss function is reduced after the first tree. It turns out that $w_1(\vx)$ is roughly negative gradient of the loss function evaluated at the initial guess, $w_1(\vx) \propto - \frac{\partial L(y)}{\partial y}|_{y(\vx)=0}$.  The second tree will fit the difference between the true value and the output from the first tree, namely, $Y(\vx)-w_1(\vx)$ by searching for the ouput $w_2(\vx)$ to maximize $L(w_1(x)) - L(w_1(x)+w_2(x))$. $w_2(\vx)$ is found to be roughly negative gradient of the loss function evaluated at $w_1(\vx)$, namely, $w_2(\vx)\propto - \frac{\partial L(y)}{\partial y}|_{y(\vx)=w_1(\vx)}$. This process can be repeated until the performance is stable.

The two methods also use different node-split schemes in building a tree. The AdaBDT relies on the Gini index (denoted by $G$). It is defined as $G \equiv Np(1-p)$ with $N$ being the number of instances in a node and $p$ is the fraction of signal instances (purity). The node split is determined by maximizing $G - G_L - G_R$, where $G$, $G_L$ and $G_R$ are the Gini indices of the mother node and two daughter nodes respectively. For the GradBDT, the node split is determined by maximizing the reduction of the loss function, namely, maximizing $L - L_L - L_R$ with $L$, $L_L$ and $L_R$ being the loss function values of the mother node and two daughter nodes respectively.

For a $m$-tree training, the final output is the combination of all outputs, namely, $y_m(\vx) = \sum_{i=1}^m k_i(\vx)\alpha_i$ for the AdaBDT and $y_m(\vx) = \sum_{i=1}^m w_i(\vx)$ for the GradBDT.

\section{Score distribution in the Adaptive BDT method}~\label{sec:AdaBDT}

Let $g_m(y_m)$ be the PDF of the BDT score $y_m$ for a training with $m$ trees. As the training is a process of iteration, we can try to find the relation between $g_m(y_m)$ and the score PDF of previous $m-1$ trees, $g_{m-1}(y_{m-1})$. As the score is additive, we have 
\begin{equation}
y_m(\vec{x}) = y_{m-1}(\vec{x}) + \alpha_m k_m(\vec{x})\:,
\end{equation}
with $k_m$ and $\alpha_m$ being the output and weight of the $m$-th tree. Ignoring the dependence upon $\vx$ and taking $y_m$, $y_{m-1}$ and $k_m$ as random variables, the PDF of $y_m$ can be obtained from the PDFs of $y_{m-1}$ and $k_m$.  
If further assuming $y_{m-1}$ and $k_m$ are independent random variables, we have
\begin{eqnarray}
\int_{y}^{y+\delta} g_m(y_m) dy_m &=& \int_{y<y_m<y+\delta} g_{m-1}(y_{m-1})f_m(k_m)dy_{m-1}dk_m\nonumber \\
&=& \int_{y<y_m<y+\delta} g_{m-1}(y_{m}-\alpha_mk_m)f_m(k_m)dy_mdk_m \:,\label{eq:AdaBDT_key}
\end{eqnarray}
where $f_m(k_m)$ is the PDF of $k_m$. Letting $\epsilon_m$ be the misclassification rate of the $m$-th tree and noting that $k_m$ has only two values, it is easy to see that 
\begin{equation}
f_m^S(1) = (1-\ep_m^S)\delta(k_m-1) + \ep_m^S\delta(k_m+1) 
\end{equation}
for signal and 
\begin{equation}
f_m^B(-1) = (1-\ep_m^B)\delta(k_m+1) + \ep_m^B\delta(k_m-1)
\end{equation}
for background. Here the subscript ``S'' ( ``B'') denotes the signal (background) category. $\delta(x)$ is the Dirac delta function. Therefore, we obtain the evolution formula of the score PDF.
\begin{eqnarray}\label{eq:AdaBDT_PDF_evolution}
 && g_m^S(y) = g_{m-1}(y-\alpha_m)(1-\epsilon_m^S) + g_{m-1}(y+\alpha_m)\epsilon_m^S\\
 && g_m^B(y) = g_{m-1}(y+\alpha_m)(1-\epsilon_m^B) + g_{m-1}(y-\alpha_m)\epsilon_m^B 
\end{eqnarray}
where the best $\alpha_m$ is found to be $\frac{1}{2}\ln\frac{1-\epsilon_m}{\epsilon_m}$~\cite{friedman1,rojas} with $\epsilon_m$ being the overall misclassification. The best $\alpha_m$ can be seen from the maximal reduction of the loss function.
\begin{eqnarray}\label{eq:AdaBDT_Lm0}
   L(y_m) &=& \sum_{\vx_i\in S} e^{-y_m(\vx_i)} + \sum_{\vx_i \in B} e^{y_m(\vx_i)} \\
   &=&\int_{-\infty}^{+\infty}e^{-y} g_m^S(y)dy + \int_{-\infty}^{+\infty} e^{y} g_m^B(y)dy 
\end{eqnarray}
Using the evolution formula in Eq.~\ref{eq:AdaBDT_PDF_evolution}, we obtain
\begin{eqnarray}\label{eq:AdaBDT_Lm}
   L(y_m) =&&(e^{-\al_m}(1-\ep_m^S) + e^{\al_m}\ep_m^S) \int_{-\infty}^{+\infty}e^{-y} g_{m-1}^S(y)dy \nonumber \\
   &&+ (e^{-\al_m}(1-\ep_m^B) + e^{\al_m}\ep_m^B) \int_{-\infty}^{+\infty} e^{y} g_{m-1}^B(y)dy 
\end{eqnarray}
Fixing $\ep_m^S$ and $\ep_m^B$, the best $\al_m$ is obtained from the condition $\partial L/\partial \alpha_m = 0$ and is written as $\frac{1}{2}\ln\frac{1-\ep_m}{\ep_m}$ with $\ep_m$ defined in the following equation.
\begin{equation}\label{eq:AdaBDT_epm}
\ep_m \equiv \frac{\ep_m^S\int_{-\infty}^{+\infty}e^{-y_{m-1}} g_{m-1}^S(y_{m-1})dy_{m-1} + \ep_m^B\int_{-\infty}^{+\infty} e^{y_{m-1}} g_{m-1}^B(y_{m-1})dy_{m-1}}{\int_{-\infty}^{+\infty}e^{-y_{m-1}} g_{m-1}^S(y_{m-1})dy_{m-1} + \int_{-\infty}^{+\infty} e^{y_{m-1}} g_{m-1}^B(y_{m-1})dy_{m-1}}
\end{equation}
Here the integration variable $y$ is written as $y_{m-1}$ for better understanding. The definition $\ep_m$ results from the boosting mechanism in the AdaBDT method. The instances misclassified by the $(m-1)$-th tree is applied a weight  of $e^{\alpha_{m-1}}$ as indicated by the factor $e^{-y_{m-1}}$ ($e^{y_{m-1}}$) for signal (background). 

In the following derivation, we only focus on the score PDF (which is actually independent upon the boosting mechanism)  for the signal and neglect the subscript ``S'' for simplicity. Here are the score PDFs for the first two trees.
\begin{eqnarray}
g_1(y) = && g_0(y-\al_1)(1-\ep_1) + g_0(y+\al_1)\ep_1 \\
g_2(y) = && g_1(y-\al_2)(1-\ep_2) + g_1(y+\al_2)\ep_2 \\
= && g_0(y-\al_1-\al_2)(1-\ep_1)(1-\ep_2) \nonumber \\
&& + g_0(y+\al_1-\al_2)\ep_1(1-\ep_2) \nonumber \\
&& + g_0(y-\al_1+\al_2)(1-\ep_1)\ep_2 \nonumber \\
&& + g_0(y+\al_1+\al_2)\ep_1\ep_2
\end{eqnarray}
For the first tree, there will be only two weighted output values for a signal instance, namely, $+\al_1$ with the probability $1-\ep_1$ and $-\al_1$ with the probability $\ep_1$. Therefore, $g_0(y)$ is simply $\delta(y)$. Furthermore, even if an instance is classified as a signal instance by the first tree, it will be classified as either a signal or a background instance by the second tree. So there will be 4  outputs in total for two trees with different probabilities. The probability is actually the product of the classification probability from each tree. In the general case, the signal PDF can be written in the following compact form.
\begin{equation}
g_m(y) = \sum_{\sg_1=\pm1}\cdots\sum_{\sg_m=\pm1}\delta(y+\sum_{i=1}^m\sg_i\al_i) \Pi_{i=1}^{m}(\frac{1-\sg_i}{2} + \sg_i\ep_i)
\end{equation}

Before calculating the final PDF, $g_m(y)$, an example is presented to illustrate how the BDT score distribution evolves with the number of trees. Here are a few items for the example.

\begin{itemize}
\item The score PDF is obtained from the iteration formula $y_m = y_{m-1} + \alpha_m k_m$.
\item Assume the misclassification rate $\ep_m^S = 0.5 -0.4\times e^{m/5}$ for the signal sample and $\ep_m^B = 0.5 - 0.3\times e^{m/5}$ for the background sample.
\item The binned significance $Z$ is calculated according to Eq.~\ref{eq:binZ} with the total number of signal (background) instances, denoted by $N_s$ ($N_b$), being 1 (100). This is used to measure the separation between signal and background with increasing number of trees.
   \begin{equation}\label{eq:binZ}
	Z \equiv \sqrt{\sum_{i=1}^{\nbins} 2\left((s_i+b_i)\ln(1+\frac{s_i}{b_i}) - s_i\right)}
   \end{equation}
   where $\nbins$ is the number of bins in the BDT score distribution and $s_i$ ($b_i$) is the number of signal (background) instances in the $i$-th bin.
\item Use a narrow Gaussian distribution with the mean 0 and the standard deviation 0.001 to imitate the delta function $\delta(y)$
\end{itemize} 

\begin{figure}[htbp] 
\includegraphics[width =0.45\textwidth]{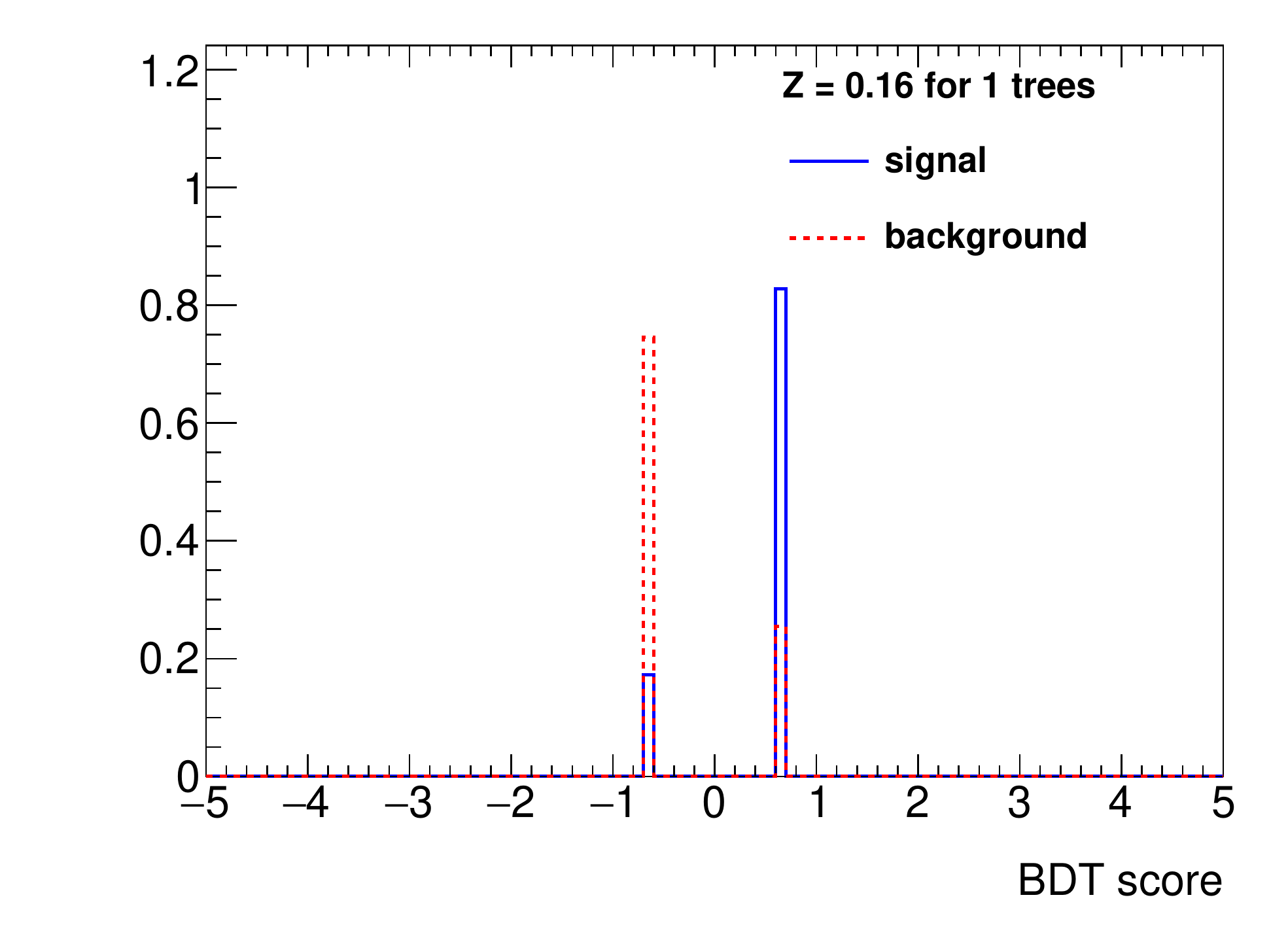} 
\includegraphics[width =0.45\textwidth]{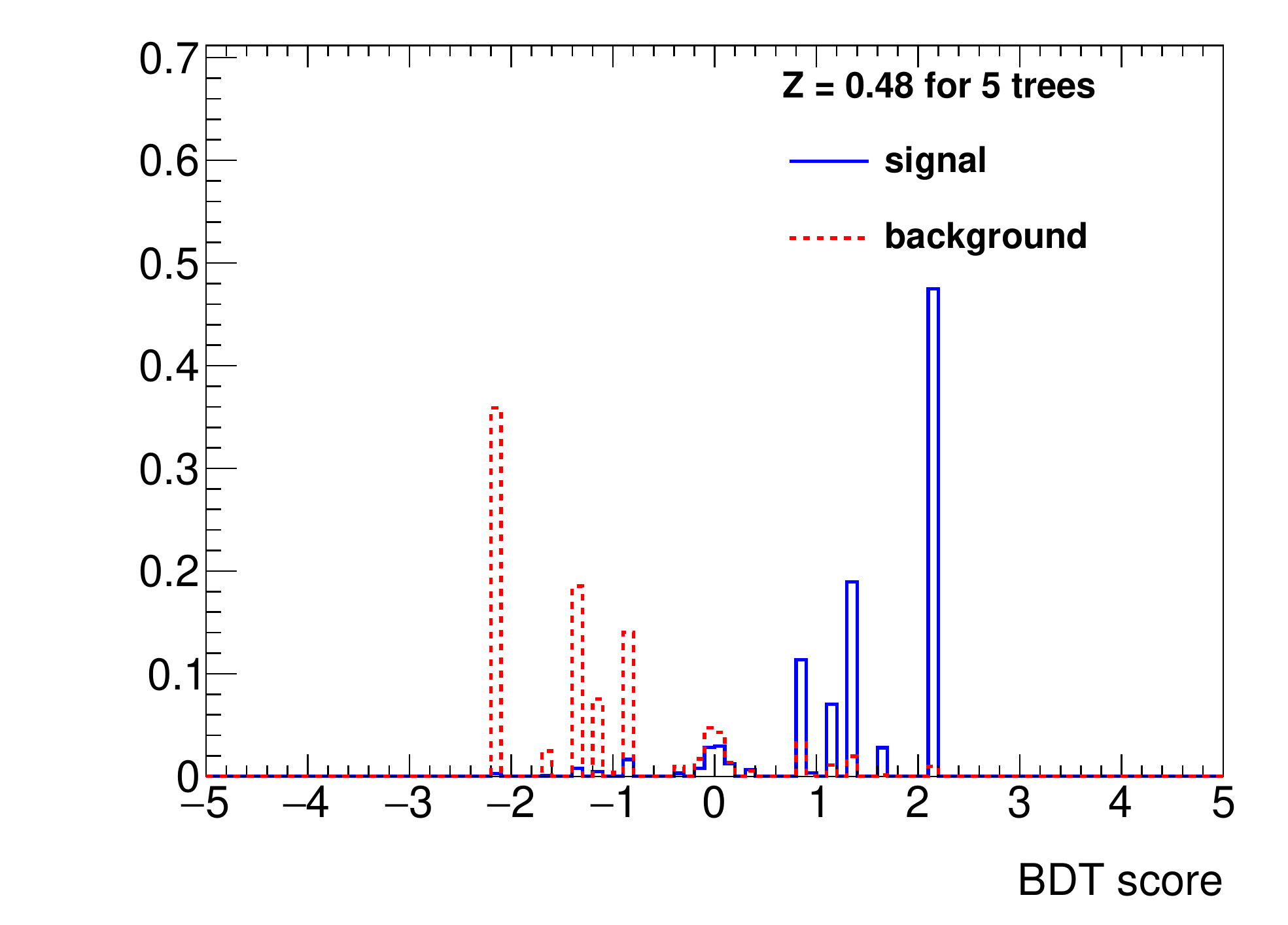} \\
\includegraphics[width =0.45\textwidth]{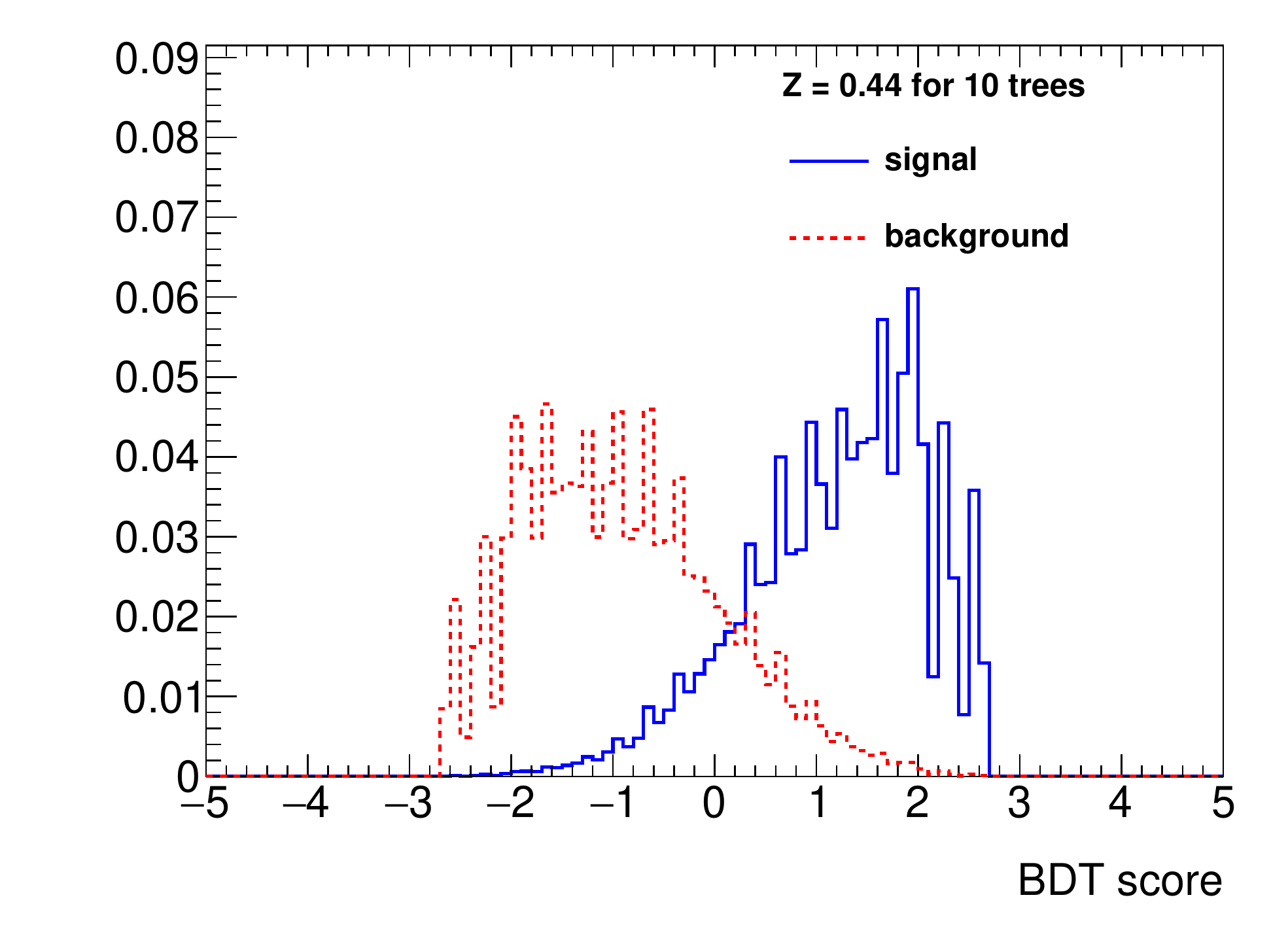} 
\includegraphics[width =0.45\textwidth]{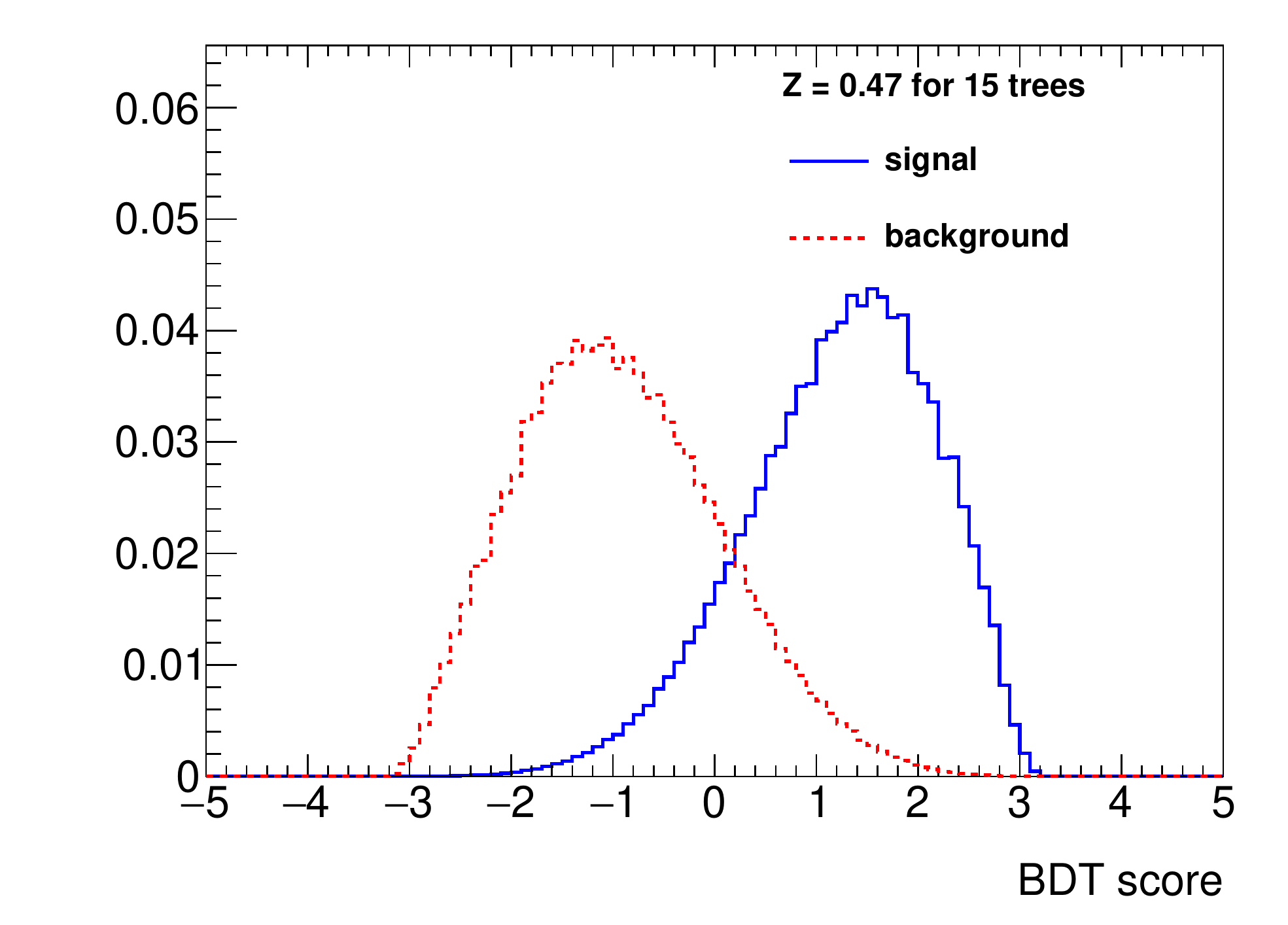}
  \caption{\label{fig:adabdt_example}
          An example to illustrate how the score probability distribution evolves with increasing number of trees. Top left: 1 tree, top right: 5 trees, bottom left: 10 trees and bottom right: 15 trees. In legend, the quantity $Z$ represents the statistical significance as explained in the text.
	        }
\end{figure}

The score PDFs for 1, 5, 10 and 15 trees are shown in Fig.~\ref{fig:adabdt_example}. We can see that the score PDFs are already stable after only 15 trees. 
The reason is that the misclassification rate $\ep^S$ ($\ep^B$) is close to 0.1 (0.2) for the first few trees and we are actually using strong classifiers to speed up the evolution. In the weak-learner limit, the misclassification rate will be close to the random-guess rate 0.5. 
We also see that the final PDFs are Gaussian-like as will be shown below.

To derive the score PDF, we resort to the characteristic function, denoted by $\phi_y(t)$. Here is the logarithmic characteristic function for the signal score PDF (see Appendix~\ref{sec:app_character} for the calculation details). Let us pick up the superscript ``S'' and ``B''.

\begin{equation}
\ln \phi_y^S(t) = \sum_{i=1}^m \ln (\cos(-t\al_i) + i(1-2\ep_i^S)\sin(t\al_i))
\end{equation}

Applying the weak-learner limits ($\ep_i^S \to 0.5^-$, $\ep_i^B \to 0.5^-$ and $\alpha_i \to 1-2\ep_i \to 0^+$) and keeping the terms up to the order of $\alpha_i^2$, the equation above becomes
\begin{equation}
   \ln\phi_y^S(t) \approx  -\frac{1}{2}(\sigma_m^S)^2t^2 + i\mu_m^S t
\end{equation}
with
\begin{eqnarray}\label{eq:adabdt_S_musigma}
   && \sigma_m^S = \sqrt{\sum_{i=1}^m \al_i^2} \:, \quad \mu_m^S = \sum_{i=1}^{m} (1-2\ep_i^S)\al_i \: .
\end{eqnarray}

Therefore, we obtain the PDF of the signal BDT score.

\begin{equation}
   g_m^S(y) \approx \frac{1}{\sqrt{2\pi}\sigma_m^S}e^{-\frac{(y-\mu_m^S)^2}{2(\sigma_m^S)^2}}
\end{equation}
It is a Gaussian distribution with the mean $\mu_m^S$ and the standard deviation $\sigma_m^S$. Similarly, the PDF of the background BDT score is
\begin{equation}
   g_m^B(y) \approx \frac{1}{\sqrt{2\pi}\sigma_m^B}e^{-\frac{(y-\mu_m^B)^2}{2(\sigma_m^B)^2}}
\end{equation}
with
\begin{eqnarray}\label{eq:adabdt_B_musigma}
   && \sigma_m^B = \sqrt{\sum_{i=1}^m \al_i^2} \:, \quad \mu_m^B = - \sum_{i=1}^{m} (1-2\ep_i^B)\al_i \: .
\end{eqnarray}

From Eq.~\ref{eq:adabdt_S_musigma} and Eq.~\ref{eq:adabdt_B_musigma}, we can see that as the number of trees increases, the distance between the two mean values is larger, and the variance of either score PDF is bigger at the same time. For the present approximation, both distributions have the same variance, $\sigma\equiv \sigma_m^S=\sigma_m^B$.

The Gaussian distribution will be corrected if we keep more terms. As shown in Appendix~\ref{sec:app_character}, the signal score PDF becomes (up to the order of $\al_i^4$) 
 
\begin{equation}
   g_m^S(y) = P(\frac{\mu_m^S-y}{\sigma_m^S}) \frac{1}{\sqrt{2\pi}\sigma_m^S} e^{-\frac{(y-\mu_m^S)^2}{2(\sigma_m^S)^2}} \:,
\end{equation}
where $P(\frac{\mu_m^S-y}{\sigma_m^S})$ is a polynomial about $\frac{\mu_m^S-y}{\sigma_m^S}$ to correct the Gaussian distribution as defined below.
\begin{equation}
   P(x) \equiv 1+\frac{3c_4}{(\sigma_m^S)^4} - 2 \frac{c_3}{(\sigma_m^S)^3}x-6\frac{c_4}{(\sigma_m^S)^4}x^2 + \frac{c_3}{(\sigma_m^S)^3}x^3 + \frac{c_4}{(\sigma_m^S)^4}x^4 \: . 
\end{equation}
with
\begin{eqnarray}
&& c_3 = \frac{1}{3} \sum_{i=1}^m (1-2\ep_i^S)\al_i^3 \:,\quad c_4 = \frac{-1}{12}\sum_{i=1}^m \al_i^4 \: ,\\
&& \sigma_m^S = \sqrt{\sum_{i=1}^m \al_i^2[1-(1-2\ep_i^S)^2]} \: .
\end{eqnarray}

It should be emphasized that the derivation of the score PDF in AdaBDT method does not rely on the details on the boosting mechanism or on the choice of loss function. In fact, we can apply the boosting mechanism of the AdaBDT method to any loss function, or inversely, find the best boosting mechanism according to any loss function.  We can take $L(y) = \sum_{\vx}\frac{1}{2}(y(\vx)-Y(\vx))^2$ as example to show it in the same way as we derive Eq.~\ref{eq:AdaBDT_Lm0} to Eq.~\ref{eq:AdaBDT_epm}. 
\begin{eqnarray}
L(y_m) =&& \sum_{\vx} \frac{1}{2}(y(\vx)-Y(\vx))^2 \\
=&& \intaz \frac{1}{2}(y-1)^2g_m^S(y)dy + \intaz \frac{1}{2}(y+1)^2g_m^B(y)dy  \\
=&& \intaz \frac{1}{2}(y-1)^2(g_{m-1}^S(y-\al_m)(1-\ep_m^S) + g_{m-1}^S(y+\al_m)\ep_m^S) \nonumber \\
 &&+\intaz \frac{1}{2}(y+1)^2(g_{m-1}^S(y+\al_m)(1-\ep_m^S) + g_{m-1}^S(y-\al_m)\ep_m^S)
\end{eqnarray}
For convenience, we define $\langle f(y)\rangle_m^S \equiv \intaz f(y) g_m^S(y)dy$. The loss function becomes
\begin{eqnarray}
L(y_m) =&&\frac{1}{2}\langle (y-1)^2\rangle_m^S + \frac{1}{2}\langle(y+1)^2\rangle_m^B \\
=&&\frac{1}{2}[ \langle (y-1+\al_m)^2 \rangle_{m-1}^S(1-\ep_m^S) + \langle (y-1-\al_m)^2\rangle_{m-1}^S \ep_m^S] \nonumber \\
&&+ \frac{1}{2}\langle (y+1-\al_m)^2 \rangle_{m-1}^B(1-\ep_m^B) + \langle (y+1+\al_m)^2\rangle_{m-1}^B \ep_m^B] \\
=&& L(y_{m-1}) + A \al_m + \al_m^2 
\end{eqnarray}
with 
\begin{equation}
A \equiv \langle y-1\rangle_{m-1}^S(1-2\ep_m^S) - \langle y+1 \rangle_{m-1}^B(1-2\ep_m^B) 
\end{equation}
The minimization condition $\partial L(y_m)/\partial \alpha_m$ gives $\alpha_m = -\frac{A}{2}$. If $\ep_m$ is defined as $\ep_m \equiv 1 + \frac{A}{2}$, namely, 
\begin{equation}\label{eq:eq:boost_1}
\ep_m \equiv \frac{1}{2}(\langle y\rangle_{m-1}^S -\langle y\rangle_{m-1}^B) + \ep_m^S\langle 1-y\rangle_{m-1}^S + \ep_m^B \langle 1+y\rangle_{m-1}^B \:,
\end{equation}
then we have $\al_m = 1- \ep_m$ and $L(y_m) = L(y_{m-1})- (1-\ep_m)^2 \leq L(y_{m-1})$. Here $\ep_m$ is apparently not the overall misclassification rate, but reflects this information. The first term in the definition of $\ep_m$ is not associated with $\ep_m^S$ or $\ep_m^B$ and is not relevant with the boosting mechanism. According to the second term $\ep_m^S \langle 1-y\rangle_{m-1}^S$, the weight for a signal instance incorrectly identified by the $m-1$-th tree will increase by a factor $\alpha_{m-1}$ because
\begin{equation}
\ep_m^S\langle 1-y_{m-1}\rangle_{m-1}^S = \ep_m^S\langle 1-y_{m-2} - k_{m-1}\alpha_{m-1}\rangle_{m-1}^S = \ep_m^S\langle 1-y_{m-2} + \alpha_{m-1}\rangle_{m-1}^S \: , 
\end{equation}
where $k_{m-1}=-1$ for a misclassified signal instance. The same weight increasement also applies to the misclassified background instances. For any loss function, the boosting mechanism can be obtained in a similar way. If it is differentiable, we can show it generally in the following way. Let $f(y(\vx),Y(\vx))$ denote any differentiable loss function for one instance. The total loss function is
\begin{equation}
L(y_m, Y) = \langle f(y, 1)\rangle_m^S + \langle f(y,-1)\rangle_m^B \: .
\end{equation}
In the weak-learner limit, $\alpha_m$ is small and we can adopt the approximation below.
\begin{eqnarray}
\langle f(y, 1)\rangle_m^{S} = && \langle f(y+\alpha_m, 1)\rangle_{m-1}^S(1-\ep_m^S) + \langle f(y-\al_m,1)\rangle_{m-1}^S \ep_m^S \\
\approx  && \langle f(y, 1) + \frac{\partial f}{\partial y}\al_m + \frac{1}{2}\frac{\partial^2f}{\partial y^2}\al_m^2\rangle_{m-1}^S(1-\ep_m^S) \nonumber \\
&&+\langle f(y, 1) - \frac{\partial f}{\partial y}\al_m + \frac{1}{2}\frac{\partial^2f}{\partial y^2}\al_m^2\rangle_{m-1}^S\ep_m^S\\
\end{eqnarray}
A similar approximation also applies to the background sample. The loss function becomes
\begin{equation}
L(y_m) = L(y_{m-1}) + A_1 \al_m + \frac{1}{2}A_2 \al_m^2 \: ,
\end{equation}
with
\begin{eqnarray}
A_1 &=& \langle \frac{\partial f}{\partial y} \rangle_{m-1}^S(1-2\ep_m^S) + \langle \frac{\partial f}{\partial y} \rangle_{m-1}^B(1-2\ep_m^B) \:, \\
A_2 &=& \langle \frac{\partial^2f}{\partial y^2} \rangle_{m-1}^S + \langle \frac{\partial^2f}{\partial y^2} \rangle_{m-1}^B \: .
\end{eqnarray}
$\partial L(y_m)/\partial \al_m=0$ gives $\al_m = \frac{-A_1}{A_2}$ and $L(y_m) = L(y_{m-1}) - \frac{A_1^2}{2A_2}$ \:. The boosting mechanism can be found from the terms related with $\ep_m^S$ and $\ep_m^B$ in the expression of $\al_m$. For signal, it is 
\begin{equation}
- \ep_m^S\langle \frac{\partial f}{\partial y} \rangle_{m-1}^S = \ep_m^S \intaz -\frac{\partial f(y,1)}{\partial y}|_{y=y_{m-1}} g_{m-1}^S(y_{m-1}) dy_{m-1} \:.
\end{equation}
Explicitly for $f(y,Y) = e^{-yY}$, the misclassified siganl instances should be applied a weigh of $e^{\al_{m-1}}$ at the $(m-1)$-th tree as indicated by
\begin{equation}
-\frac{f(y,1)}{\partial y}|_{y=y_{m-1}} = e^{-y_{m-1}} = e^{-y_{m-2}}e^{\al_{m-1}} \: .
\end{equation}
From the derivation above, we have already seen the similarity between the boosting mechanism in AdaBDT and that in the GradBDT presented in next section.

\section{Score distribution in the Gradient BDT method}~\label{sec:GradBDT}
To derive the Gradient BDT score distribution, it is necessary to review some details. Here we follow the ideas of the XGBoost algorithm~\cite{xgboost}, which is developed from the original method~\cite{friedman2}. Suppose we have $m$ trees, the BDT score is

\begin{equation}\label{eq:GradBDT_iteration}
y_m(\vx_i) = y_{m-1}(\vx_i) + w_m(\vx_i) \: .
\end{equation}
Taking $w_m(\vx_i)$ as a small quantity and expanding the loss fuction around $y_{m-1}(\vx_i)$ to the order of $w_m^2$, we have
\begin{eqnarray}
L_m \equiv &&\sum_{\vx_i} l(y_m(\vx_i)) = \sum_{\vx_i}l(y_{m-1}(\vx_i)+w_m(\vx_i)) \\
\approx && \sum_{\vx_i} l(y_{m-1}(\vx_i)) + d_{m-1}(\vx_i) w_m(\vx_i) + \frac{1}{2}h_{m-1}(\vx_i)w_m^2(\vx_i)\\
\end{eqnarray}
with
\begin{equation}\label{eq:dh}
d_{m-1} \equiv \frac{\partial l(y)}{\partial y}|_{y=y_{m-1}} \:, \quad h_{m-1} \equiv \frac{\partial^2 l(y)}{\partial y^2}|_{y=y_{m-1}} \: .
\end{equation}

In practice, each tree will only have limited number of terminal nodes (denoted by $J$). The instances falling into the same terminal node (denoted by $R_j, j=1,2,\cdots,J$) will be given the same tree output, $w_m(R_j)$. The loss function then becomes
\begin{eqnarray}\label{eq:w}
L_m \approx && \sum_{\vx_i} l(y_{m-1}(\vx_i)) + d_{m-1}(\vx_i)w_m(\vx_i) + \frac{1}{2}h_{m-1}(\vx_i)w_m^2(\vx_i)\\
= && L_{m-1}+\sum_{j=1}^{J} \left(\sum_{\vx_i \in R_j} d_{m-1}(\vx_i)\right) w_m(R_j) + \left(\sum_{\vx_i\in R_j}h_{m-1}(\vx_i)\right)w_m^2(R_j)
\end{eqnarray}

Minimizing the loss function gives
\begin{equation}
w_m(R_j) = - \frac{\sum_{\vx_i\in R_j} d_{m-1}(\vx_i)}{\sum_{\vx_i\in R_j}h_{m-1}(\vx_i)} \: .
\end{equation}
and the reduction of the loss function due to the $m$-th tree is
\begin{equation}\label{eq:GradBDT_split}
\Delta L_m \equiv L_m - L_{m-1} = - \frac{1}{2}\sum_{j=1}^J \frac{(\sum_{\vx_i\in R_j}d_{m-1}(\vx_i))^2}{\sum_{\vx_i \in R_j}h_{m-1}(\vx_i)} \: .
\end{equation}
This is used to determine which variable should be used to split a node and also the best splitting position. Let $R$ denote a node and $R_l$ and $R_r$ denote its daughter nodes corresponding to the requirement $x<c$ and $x>c$ respectively, where $x$ is one of the variables in training and $c$ is a possible split postion. $x$ and $c$ are chosen so as to maximize the reduction of the loss function, namely, 
\begin{equation}\label{eq:GradBDT_split1}
\frac{1}{2}\frac{(\sum_{\vx_i \in R_l}g(\vx_i))^2}{\sum_{\vx_i\in R_l}h(\vx_i)} + \frac{1}{2}\frac{(\sum_{\vx_i \in R_r}g(\vx_i))^2}{\sum_{\vx_i\in R_r}h(\vx_i)}-\frac{1}{2}\frac{(\sum_{\vx_i \in R}g(\vx_i))^2}{\sum_{\vx_i\in R}h(\vx_i)} \: .
\end{equation}

After reviewing these basic ideas, we start with the BDT score iteration relation from $(m-1)$-th tree to the $m$-th tree.
\begin{eqnarray}
&& y_m(\vx_i) = y_{m-1}(\vx_i) + \sum_{j=1}^{J} \delta_{\vx_i,R_j} w_m(R_j)
\end{eqnarray}
Here $\delta_{\vx_i, R_j}$ is 1 if $\vx_i$ falls into the node $R_j$ and 0 otherwise. Note that all terminal nodes $R_j$ do not overlap. For simplicity, we use the loss fucntion $l(y_m(\vx_i)) = \frac{1}{2}(y_m(\vx_i)-Y(\vx_i))^2$. 
According to Eqs.~\ref{eq:dh} and ~\ref{eq:w}, we have 

\begin{equation}
y_m(\vx_i) = y_{m-1}(\vx_i) - \sum_{j=1}^{J} \delta_{\vx_i,R_j} \frac{1}{N_{R_j}}\left(\sum_{\vx_i \in R_j} y_{m-1}(\vx_i) - Y(\vx_i)\right) \: ,
\end{equation}
where $N_{X}$ denotes the number of instances in the region $X$.

Let $p_{m,R_j}$ denote the signal fraction in the node $R_j$ (thus the background fraction is $1-p_{m,R_j}$). Let $f_{m,R_j}$ denote the fraction of total number of instances in the node $R_j$, namely, $N_{R_j}/\sum_{j=1}^JN_{R_j}$. Let $S\cap R_j$ and $B\cap R_j$ denote the set of signal and background instances in the node $R_j$. Then we have $p_{m,R_j} = N_{S\cap R_j}/N_{R_j}$, $\sum_j f_{m,R_j} = 1$, $\sum_j f_{m,R_j}p_{m,R_j} = \frac{1}{2}$ and $\sum_j f_{m,R_j}(1-p_{m,R_j}) = \frac{1}{2}$ ( this is because both signal and background samples are renormalized to be 1 instance by definition ). The output for the $R_j$ at the $m$-th tree becomes

\begin{eqnarray}
   -w_m(R_j) &=& \frac{1}{N_{R_j}}\left(\sum_{\vx_i \in R_j} y_{m-1}(\vx_i) - Y(\vx_i)\right) \\
&=&\frac{1}{N_{R_j}} \left( N_{S\cap R_j} \frac{\sum_{\vx_i \in S\cap R_j}y_{m-1}(\vx_i)-1}{N_{S\cap R_j}} + N_{B\cap R_j} \frac{\sum_{\vx_i \in B\cap R_j}y_{m-1}(\vx_i)+1}{N_{B\cap R_j}}\right) \\
&=&  p_{m,R_j}(z_{m-1,R_j}^S-1) + (1-p_{m,R_j})(z_{m-1,R_j}^B+1) \:,
\end{eqnarray}
where
\begin{equation}
z_{m-1,R_j}^S \equiv \frac{\sum_{\vx_i \in S\cap R_j}y_{m-1}(\vx_i)}{N_{S\cap R_j}} \:, \quad  z_{m-1,R_j}^B \equiv \frac{\sum_{\vx_i \in B\cap R_j}y_{m-1}(\vx_i)}{N_{B\cap R_j}} \:.
\end{equation}

Let $\mu_{m-1}$ and $\sigma_{m-1}^2$ denote the expectation value and variance of the distribution of $y_{m-1}$. We assume they exist and assume the Central Limit Theorem (CLT) applies here. $z_{m-1,R_j}$ will abide by a Gaussian distribution (letting $G(x|\mu,\sigma) \equiv \frac{1}{\sqrt{2\pi}\sigma}e^{-\frac{(x-\mu)^2}{2\sigma^2}}$).
\begin{equation}
z_{m-1,R_j}^S \sim G(\mu_{m-1}^S, \frac{\sigma_{m-1}^S}{\sqrt{N_{S\cap R_j}}})\:, \quad
z_{m-1,R_j}^B \sim G(\mu_{m-1}^B, \frac{\sigma_{m-1}^B}{\sqrt{N_{B\cap R_j}}})
\end{equation}
where the superscripts ``S'' and ``B'' denote signal and background respectively as before. In the limit of big sample size, the distribution of $z_{m-1,R_j}^{S/B}$ is very peaky around the mean value $\mu_{m-1}^{S/B}$ and we can replace $z_{m-1,R_j}^{S/B}$ by its mean value approximately. Then the output for the instance $\vx_i$ from the $m$-th tree becomes
\begin{eqnarray}
w_m(\vx_i)&\approx&-\sum_{j=1}^{J} \delta_{\vx_i, R_j} \left[ p_{m,R_j}(\mu_{m-1}^S-1) + (1-p_{m,R_j})(\mu_{m-1}^B+1)\right]
\end{eqnarray}

To connect the score PDFs of $y_m$ and $y_{m-1}$, we need to know the possible values for $w_m$ and the corresponding probilities. The probability of a signal instance falling to the node $R_j$ should be proportional to the fraction of signal instances in $R_j$. This is $f_{m,R_j}p_{m,R_j}/\sum_{i=1}^Jf_{m,R_i}p_{m,R_i} = 2f_{m,R_j}p_{m,R_j}$ and similar argument applies to background. Let us only consider two nodes, namely, $J=2$. $w_m(\vx_i)$ takes only two possible values, namely
\begin{eqnarray}
   && -w_m(\vx_i) \nonumber \\
=&& \delta_{\vx_i, R_1} \left[ p_{m,R_1}(\mu_{m-1}^S-1) + (1-p_{m,R_1})(\mu_{m-1}^B+1)\right]\\
&&+ \delta_{\vx_i, R_2} \left[ \frac{1-2f_{m,R_1}p_{m,R_1}}{2(1-f_{m,R_1})}(\mu_{m-1}^S-1) + \frac{1-2f_{m,R_1}+2f_{m,R_1}p_{m,R_1}}{2(1-f_{m,R_1})}(\mu_{m-1}^B+1)\right] \:,
\end{eqnarray}
where $p_{m,R_2} = \frac{1-2f_{m,R_1}p_{m,R_1}}{2(1-f_{m,R_1})}$ is used. The corresponding probability for  a signal instance is
\begin{eqnarray}
&& \text{Prob}(R_1) = 2 f_{m,R_1} p_{m,R_1} \\
&& \text{Prob}(R_2) = 1- 2 f_{m,R_1} p_{m,R_1}  \: .
\end{eqnarray}
In the case of $J=2$, we can drop the subscripts for convenience and define $f\equiv f_{m,R_1}$ and $p\equiv p_{m,R_1}$. The output for the $m$-th tree becomes
\begin{equation}\label{eq:w_fp}
-w_m(\vx_i)\approx \delta_{\vx_i, R_1} (2p-1)(\mu^S-1) + \delta_{\vx_i, R_2} \frac{-f}{1-f}(2p-1)(\mu^S-1)\:,
\end{equation}
where $\mu^S \approx -\mu^B$ is used. If the initial guess is 0 for all instances, this approximation is expected as signal and background play an equal role. In the equation above, $f$ and $p$ are not independent and are related by the splitting criteria shown in Eq.~\ref{eq:GradBDT_split} and Eq.~\ref{eq:GradBDT_split1}. We can take $p$ as a function of $f$ and $p(1)=\frac{1}{2}$ (if a node has all the instances, then the signal fraction in that node is $\frac{1}{2}$ due to the initial renormalization). As all trees are weak learners, we expect that $p$ should be around $\frac{1}{2}$ and has little dependence upon $f$. According to Eq.~\ref{eq:GradBDT_split} and expanding $p(f) \approx p(1) + \frac{dp}{df}(f-1)  = \frac{1}{2} + \frac{dp}{df}(f-1)$, we have
\begin{eqnarray}
&&\frac{1}{2}f w_m(R_1)^2 + \frac{1}{2}(1-f) w_m(R_2)^2 \\
\approx && \frac{1}{2}\frac{f}{1-f}(2p-1)^2 (\mu^S - 1)^2 \\
\approx && \frac{1}{2}\left(\frac{dp}{df}\right)^2 f(1-f) (\mu^S - 1)^2 \: .
\end{eqnarray}
Maximizing the quantity above gives $f \approx \frac{1}{2}$, consistent with the intuitive picture about weak-leaner method.

Let us recover the subscripts and summarize the possible values of $w_m$ and the corresponding probabilities. For a signal instance, we have
\begin{eqnarray}
&& w_m(\vx_i) = \left\{ 
\begin{matrix}
-(2p_{m,R_1}-1)(\mu_{m-1}^S-1) & \text{Prob}( R_1)= p_{m,R_1} \\
(2p_{m,R_1}-1)(\mu_{m-1}^S-1) & \text{Prob}(R_2)= 1- p_{m,R_1}\\
\end{matrix} \: .
\right.
\end{eqnarray}
Similarly for a background instance, we have
\begin{eqnarray}
&& w_m(\vx_i) = \left\{ \begin{matrix}
(2p_{m,R_1}-1)(\mu_{m-1}^B+1) & \text{Prob}(R_1)= 1-p_{m,R_1} \\
-(2p_{m,R_1}-1)(\mu_{m-1}^B+1) & \text{Prob}(R_2)= p_{m,R_1} \\
\end{matrix} \: .
\right.
\end{eqnarray}

Taking $y_{m-1}$ and $w_{m}$ as random variables and assuming they are independent as in Sec.~\ref{sec:AdaBDT}, the BDT score PDF satisfies the following iteration relation.
\begin{eqnarray}
&&\int_{y}^{y+\delta} g_m(y_m) dy_m = \int_{y<y_m<y+\delta} g_{m-1}(y_{m-1}) f(w_{m}) dy_{m-1}dw_{m} \\
= && \int_{y}^{y+\delta} dy_m\left[g_{m-1}(y_m-w_m(R_1)) \text{Prob}(R_1) + g_{m-1}(y_m-w_m(R_2)) \text{Prob}(R_2) \right]
\end{eqnarray}
For signal, it gives
\begin{equation}\label{eq:GradBDT_evolution}
g_m^S(y) = g_{m-1}^S(y + (2p_{m,R_1}-1)(\mu_{m-1}^S-1)) p_{m,R_1} + g_{m-1}^S(y - (2p_{m,R_1}-1)(\mu_{m-1}^S-1)) (1-p_{m,R_1}) \: .
\end{equation}
Compared to the evolution formula below (copied from Eq.~\ref{eq:AdaBDT_PDF_evolution}) in the Adaptive BDT method,
\begin{equation}
g_m^S(y) = g_{m-1}^S(y-\alpha_m)(1-\epsilon_m^S) + g_{m-1}(y+\alpha_m)\epsilon_m^S
\end{equation}
they are equivalent with the following correspondence raltions
\begin{eqnarray}
&& p_{m,R_1} = \epsilon_m^S \:, \\
&& (2p_{m,R_1}-1)(\mu_{m-1}^S-1) = \alpha_m \: .
\end{eqnarray}

This observation shows the equivalence between the AdaBDT and the GradBDT with only 2-node trees. Therefore, the score PDFs in GradBDT are also approximately Gaussian functions. Based on the definitions of the expectation and variance of a PDF, 
\begin{eqnarray}
&&\mu_m = \int_{-\infty}^{+\infty} y g_m(y)dy \:, \\
   && (\sigma_m)^2 = \int_{-\infty}^{+\infty}(y-\mu_m)^2g_m(y) dy \:,
\end{eqnarray}
and using the evolution formula in Eq.~\ref{eq:GradBDT_evolution}, we obtain (some calculation details can be found in Appendix~\ref{sec:app_grad}.)
\begin{eqnarray}
   \mu_m^S &\approx& - \mu_m^B \approx 1 - \Pi_{i=1}^m 4p_{i,R_1}(1-p_{i,R_1})  \\ 
   (\sigma_m^S)^2 &\approx& (\sigma_m^B)^2 \approx \sum_{i=1}^m  (2p_{i,R_1}-1)^2 \: .
\end{eqnarray}
Noting that $4p_{i,R_1}(1-p_{i,R_1}) \leq 1$ (the equal sign holds only if $p_{i,R_1}=\frac{1}{2}$. If it happens, the training will stop because the loss function cannot be reduced further.), we expect that $\mu_m^S \to 1^-$ and $\mu_m^B \to (-1)^+$ with increasing number of trees. We do not see this behaviour in the AdaBDT method. It should be noted that we do not use the GradBDT score, $y$, directly in the official multi-variable analysis tool TMVA~\cite{TMVA}. Instead, $\tanh(y)$ is used as the final score as it maps $(-\infty, +\infty)$ to a bounded region $(-1, +1)$.

\section{Loss functions and statitical significance}\label{sec:HEP}
In many HEP analyses, people care about the sensitivity to probe rare signals with the presence of hugh background. Quantitatively, this is described by the statistical significance introduced in Eq.~\ref{eq:binZ}. In this section, let us investigate the relation between the improvement of statistical significance and the reduction of the loss function.

In the case of small signal ($N_s << N_b$), we have (some calculation details can be found in Appendix~\ref{sec:app_sig}) 
   \begin{eqnarray}
	Z^2 &\approx& \frac{N_s^2}{N_b} e^{\frac{(\mu_m^S-\mu_m^B)^2}{\sigma_m^2}} \\
	    &\approx& \frac{N_s^2}{N_b}\times\left\{ \begin{matrix}
	e^{2(\mu_m^S-\mu_m^B)} & \text{for AdaBDT} \\
	e^{4(\frac{\mu_m^S}{\sigma_m})^2} & \text{for GradBDT}\\
   \end{matrix} \right. \: .
   \end{eqnarray}

In the AdaBDT method, the loss function is $L(y) = \sum_{\vx_i} e^{-y(\vx_i)Y(\vx_i)}$. According to Eq.~\ref{eq:AdaBDT_Lm} and Eq.~\ref{eq:AdaBDT_epm}, we can obtain
\begin{equation}
L(y_m) = (e^{-\al_m}(1-\ep_m) + e^{\al_m}\ep_m)L(y_{m-1}) = 2\sqrt{\ep_m(1-\ep_m)}L(y_{m-1}) \: ,
\end{equation}
and thus $L(y_m) \leq L(y_{m-1})$. We can look at it in a different way. 
\begin{eqnarray}
   L(y_m)&=&\int_{-\infty}^{+\infty}e^{-y} g_m^S(y)dy + \int_{-\infty}^{+\infty} e^{y} g_m^B(y)dy \\
   &\approx& e^{\frac{\sigma_m^2}{2}}(e^{-\mu_m^S} + e^{+\mu_m^B}) \\
   &\approx& e^{-\frac{\mu_m^S}{2}} + e^{+\frac{\mu_m^B}{2}}\:. 
\end{eqnarray}
Noting that $\mu_m^S$ and $-\mu_m^B$ increases with $m$, we see the loss function reduces with more trees as 
\begin{equation}
   \frac{\partial L}{\partial \mu_m^S} < 0 \:, \frac{\partial L}{\partial (-\mu_m^B)} <0 \:,
\end{equation}
and the significance increases as
\begin{equation}
   \frac{\partial Z^2}{\partial \mu_m^S} > 0 \:, \frac{\partial Z^2}{\partial (-\mu_m^B)} >0 \:.
\end{equation}
Especially if the misclassification rate is similar for signal and background, we have $\mu_m^S \approx -\mu_m^B$ and 
\begin{equation}
   Z^2 \approx \frac{1}{4}\frac{N_s^2}{N_b}L^{-2} \: .
\end{equation}
It shows that the minization of the loss function is equivalent to the maximization of the statistical significance in this sense.

For the Graident BDT, the loss function $L(y) = \sum_{\vx_i}\frac{1}{2}(y(\vx_i)-Y(\vx_i))^2$ is used in last section. If $m$ trees are used in the training, it is
\begin{eqnarray}
   L(y_m) &=& \sum_{\vx_i\in S}\frac{1}{2}(y_m(\vx_i)-1)^2 + \sum_{\vx_i\in B} \frac{1}{2}(y_m(\vx_i)+1)^2\\
&=&\int_{-\infty}^{+\infty}\frac{1}{2}(y-1)^2 g_m^S(y)dy + \int_{-\infty}^{+\infty} \frac{1}{2} (y+1)^2 g_m^B(y)dy \\
&=& \frac{1}{2}[(\sigma_m^S)^2 + (\mu_m^S-1)^2 + (\sigma_m^B)^2 + (\mu_m^B+1)^2]\\
&\approx& \sigma_m^2 + (\mu_m^S -1)^2
\end{eqnarray}
Using the evolution relation of $\mu_m^S$ and $\sigma_m$ with $m$ shown in Eq.~\ref{eq:GradBDT_mu_evolve} and Eq.~\ref{eq:GradBDT_sigma_evolve}, we can expand $L(y_m)$ and $Z^2$ to the order of $(2p_{i,R_1}-1)^2$ and obtain
\begin{eqnarray}
   && L(y_m) \approx L(y_{m-1}) - (2p_{m,R_1}-1)^2(\mu_{m-1}^S-1)^2 < L(y_{m-1}) \\
   && Z^2(\mu_m^S,\sigma_m) \approx [Z^2(\mu_{m-1}^S, \sigma_{m-1})]^{1+(2p_{m,R_1}-1)^2(1-\mu_{m-1}^S)^2} > Z^2(\mu_{m-1}^S, \sigma_{m-1}) \: .
\end{eqnarray}
This shows that the GradBDT algorithm reduces the loss function and increase the statistical significance in a step-wise way, and they are closely related by 
\begin{equation}
\frac{\ln Z^2(\mu_m^S,\sigma_m)}{\ln Z^2(\mu_{m-1}^S,\sigma_{m-1})} \approx 1 + (L(y_{m-1}) - L(y_{m}))  \: .
\end{equation}

\section{Summary}~\label{sec:summary}
 
In summary, two popular BDT algorithms, AdaBDT and GradBDT, are studied assuming all decision trees are weak learners. The formulae describing the evolution of the score and the score PDF with the number of trees are derived. They are important for further studies. The score PDF turns out to be Gaussian approximately for both method. As more trees are used in the training, the distance between the expectation value of the signal score PDF and that of the background score PDF is larger while the variance is also greater at the same time. Extension the boosting idea in the AdaBDT method to any loss function is shown to be possible. An equivalence relation is also built for the GradBDT with 2-node trees and the AdaBDT. In addition, for the applications in HEP, we find that the improvement of the statistical significance is closely related with the reduction of the loss functions.

\section{Acknowledgement}
I would like to thank Fang Dai for encouraging words.

\appendix
\section{The characteristic function of the score PDF in the AdaBDT method}~\label{sec:app_character}
The characteristic function of the signal score PDF  for the AdaBDT method is
\begin{eqnarray}
   \phi_y^S(t) &=& \int_{-\infty}^{+\infty} g_m(y) e^{iyt} dy \label{eq:appA1}\\
			     &=& \sum_{\sg_1=\pm1}\cdots\sum_{\sg_m=\pm1}e^{-it\sum_{i=1}^m\sg_i\al_i} \Pi_{i=1}^{m}(\frac{1-\sg_i}{2} + \sg_i\ep_i^S) \label{eq:appA2}\\
		 &=& \sum_{\sg_1=\pm1}\cdots\sum_{\sg_m=\pm1}\Pi_{i=1}^m e^{-it\sg_i\al_i} (\frac{1-\sg_i}{2} + \sg_i\ep_i^S) \label{eq:appA3}\\
   &=& \Pi_{i=1}^m \sum_{\sg_i=\pm1} e^{-it\sg_i\al_i} (\frac{1-\sg_i}{2} + \sg_i\ep_i^S) \label{eq:appA4}\\
   &=& \Pi_{i=1}^m (\ep_i^S e^{-it\al_i} + (1-\ep_i^S)e^{it\al_i}) \label{eq:appA5}\\
   &=& \Pi_{i=1}^m (\cos(-t\al_i) + i(1-2\ep_i^S)\sin(t\al_i)) \label{eq:appA6}
\end{eqnarray}
The integration on $y$ is done using the property of $\delta$ function in Eq.~\ref{eq:appA2}. From Eq.~\ref{eq:appA3} to Eq.~\ref{eq:appA4}, the order of summation and product is exchanged. Then the logarithmic characteristic function becomes
\begin{equation}
\ln \phi_y^S(t) = \sum_{i=1}^m \ln(\cos(-t\al_i) + i(1-2\ep_i^S)\sin(t\al_i)) \: .
\end{equation}
With the weak-learner limit, namely, $(1-2\ep_i^S) \to 0^+$ and $\alpha \to 1- 2\ep_i \to 0^+$, and keeping the terms up to the order of $\alpha^4$, we have 
\begin{eqnarray}
   && \cos(-t\al_i) = 1 - \frac{(t\al_i)^2}{2!} + \frac{(t\al_i)^4}{4!} + \cdots \\
   && \sin(t\al_i) = t\al_i - \frac{(t\al_i)^3}{3!} + \cdots \\
   && \ln(1+x) = x - \frac{1}{2}x^2 + \cdots 
\end{eqnarray}
and 
\begin{equation}
   \ln \phi_y^S(t) \approx i\mu_m^St- \frac{1}{2}(\sigma_m^S)^2 t^2 + ic_3t^3 + c_4t^4\:,
\end{equation}
where
\begin{eqnarray}
   && \mu_m^S = \sum_{i=1}^{m} (1-2\ep_i^S)\al_i \:, \quad \sigma_m^S = \sqrt{\sum_{i=1}^m \al_i^2[1-(1-2\ep_i^S)^2]}\:, \\
   && c_3 = \frac{1}{3} \sum_{i=1}^m (1-2\ep_i^S)\al_i^3 \:, \quad c_4 = \frac{-1}{12}\sum_{i=1}^m \al_i^4 \:.
\end{eqnarray}
The BDT score PDF $g_m^S(y)$ can be obtained by the inverse fourier transformation.
\begin{eqnarray}
   g_m^S(y) &=& \frac{1}{2\pi}\int_{-\infty}^{+\infty} \phi_y^S(t) dt \\
				   &\approx& \frac{1}{2\pi}\int_{-\infty}^{+\infty} e^{i\mu_m^St- \frac{1}{2}(\sigma_m^S)^2 t^2 + ic_3t^3 + c_4t^4 } dt \\
				   &\approx& \frac{1}{2\pi}\int_{-\infty}^{+\infty} e^{i\mu_m^St- \frac{1}{2}(\sigma_m^S)^2 t^2}(1 + ic_3t^3 + c_4t^4 ) dt 
\end{eqnarray}
Using the property of the Gamma function $(\frac{\sqrt{2}}{\sigma})^{2k+1}\Gamma(\frac{2k+1}{2}) = \int_{-\infty}^{+\infty} x^{2k}e^{-\frac{1}{2}\sigma^2 x^2}dx$, it becomes
\begin{equation}
   g_m^S(y) \approx P(\frac{\mu_m^S-y}{\sigma_m^S}) \frac{1}{\sqrt{2\pi}\sigma_m^S} e^{-\frac{(y-\mu_m^S)^2}{2(\sigma_m^S)^2}} \:,
\end{equation}
where $P(\frac{\mu_m^S-y}{\sigma_m^S})$ is a polynomial about $\frac{\mu_m^S-y}{\sigma_m^S}$ to correct the Gaussian distribution and
\begin{equation}
   P(x) = 1+\frac{3c_4}{(\sigma_m^S)^4} - 2 \frac{c_3}{(\sigma_m^S)^3}x-6\frac{c_4}{(\sigma_m^S)^4}x^2 + \frac{c_3}{(\sigma_m^S)^3}x^3 + \frac{c_4}{(\sigma_m^S)^4}x^4 \: . 
\end{equation}

\section{Expectation and variance of the GradBDT score PDF}\label{sec:app_grad}
From the definition of the expectation value of a PDF,
\begin{eqnarray}
\mu_m = \int_{-\infty}^{+\infty} y g_m(y)dy
\end{eqnarray}
and using the evolution formula in Eq.~\ref{eq:GradBDT_evolution}, we obtain
\begin{eqnarray}
\mu_m^S &=& \mu_{m-1}^S - (2p_{m,R_1}-1)^2(\mu_{m-1}^S -1 ) \\
\mu_m^S-1 &=& \mu_{m-1}^S-1 - (2p_{m,R_1}-1)^2(\mu_{m-1}^S -1 ) \\
\mu_m^S-1 &=&[1- (2p_{m,R_1}-1)^2](\mu_{m-1}^S -1 ) \\
   \mu_m^S-1 &=&4p_{m,R_1}(1-p_{m,R_1})(\mu_{m-1}^S -1 ) \label{eq:GradBDT_mu_evolve}\\
		   &=& [\Pi_{i=1}^m 4p_{i,R_1}(1-p_{i,R_1})] (\mu_0 - 1) \:,
\end{eqnarray}
where $\mu_0$ is the initial guess for all instances. Therefore, we have
\begin{eqnarray}\label{eq:GradBDT_mu_S}
   \mu_m^S &=& 1 + [\Pi_{i=1}^m 4p_{i,R_1}(1-p_{i,R_1})] (\mu_0 - 1) \: . 
\end{eqnarray}

From the definition of the variance of a PDF,
\begin{equation}
(\sigma_m^S)^2 = \int_{-\infty}^{+\infty}(y-\mu_m^S)^2g_m(y) dy
\end{equation}
and using the evolution formula in Eq.~\ref{eq:GradBDT_evolution}, we obtain
\begin{eqnarray}
   (\sigma_m^S)^2 &=& (\sigma_{m-1}^S)^2 + 4(p_{m,R_1}-1)^2(2p_{m,R_1}-1)^2(\mu_{m-1}^S-1)^2 \label{eq:GradBDT_sigma_evolve}\\
			&=& \sum_{i=1}^m 4(p_{i,R_1}-1)^2 (2p_{i,R_1}-1)^2(\mu_{i-1}^S-1)^2 \\
   &=& \sum_{i=1}^m 4(p_{i,R_1}-1)^2 (2p_{i,R_1}-1)^2 ([\Pi_{j=1}^{i-1}4p_{j,R_1}(1-p_{j,R_1})](\mu_0-1))^2 \:.
\end{eqnarray}
Similarly, the expectation and variance of the background score PDF are
\begin{eqnarray}
   \mu_m^B &=& -1 + \Pi_{i=1}^m 4p_{i,R_1}(1-p_{i,R_1}) (\mu_0 +1) \label{eq:GradBDT_mu_B}\\
   (\sigma_m^B)^2&=& \sum_{i=1}^m 4p_{i,R_1}^2 (2p_{i,R_1}-1)^2 ([\Pi_{j=1}^{i-1}4p_{j,R_1}(1-p_{j,R_1})](\mu_0-1))^2 \:.
\end{eqnarray}
In the derivation, we have argued that the signal and background play equal role in the classification problem and used the approximation $\mu_m^B \approx -\mu_m^S$. From Eq.~\ref{eq:GradBDT_mu_S} and Eq.~\ref{eq:GradBDT_mu_B}, it requires that $\mu_0 \approx 0$. It means the initial guess should have litte preference. Under the weaker-learner approximation, $p_{m,R_i} \to \frac{1}{2}$ and choosing $\mu_0=0$, the expressions above can be simplified to be (keeping terms of the lowest order)
\begin{eqnarray}
   \mu_m^S &\approx& - \mu_m^B \approx 1 - \Pi_{i=1}^m 4p_{i,R_1}(1-p_{i,R_1})  \\ 
				  &=& 1 - \Pi_{i=1}^m (1- (2p_{i,R_1}-1)^2)  \\ 
   (\sigma_m^S)^2 &\approx& (\sigma_m^B)^2 \approx \sum_{i=1}^m  (2p_{i,R_1}-1)^2 \: .
\end{eqnarray}

\section{About the statistical significance}\label{sec:app_sig}
For the a binned score distribution with the bin width $\Delta y$, the statistical significance is defined in Eq.~\ref{eq:binZ}. If the signal strength is much smaller than the background ( this is the case where the HEP scientists usually apply ML methods ), the calculation can be simplified in the continuum limit.
\begin{eqnarray}
   Z^2 &=&\sum_{i=1}^{\nbins}2((s_i+b_i)\ln(1+\frac{s_i}{b_i}) - s_i) \\
	 &\approx& \sum_{i=1}^{\nbins} \frac{s_i^2}{b_i} \\
  &=& \frac{N_s^2}{N_b}\sum_{i=1}^{\nbins}\frac{(g^S(y_i))^2}{g^B(y_i)}\Delta y \\
  &\approx&\frac{N_s^2}{N_b}\int_{-\infty}^{+\infty}\frac{(g^S(y))^2}{g^B(y)}dy\\ 
  &=& \frac{N_s^2}{N_b}e^{\frac{(\mu^S-\mu^B)^2}{\sigma^2}} \: .
\end{eqnarray}
where we have assumed $s_i<<b_i$.

In the AdaBDT method, we have $\mu_m^S-\mu_m^B = \sum_{i=1}^m 2(1-2\ep_i)\al_i\approx 2\sigma_m^2$ and thus $Z^2 \approx e^{2(\mu_m^S-\mu_m^B)}$. In the GradBDT method, we have $\mu_m^S\approx - \mu_m^B$ and thus $Z^2 \approx e^{4(\mu_m^S/\sigma_m)^2}$.  

\end{document}